\documentclass[final,5p,times,preprint]{elsarticle}
\usepackage[utf8]{inputenc}
\usepackage{amsfonts}
\usepackage{amsmath}
\usepackage{amssymb}
\usepackage{bm}
\usepackage{lineno,hyperref}
\usepackage{comment}
\usepackage{xcolor}
\usepackage{booktabs}
\modulolinenumbers[5]
\usepackage[normalem]{ulem}
\usepackage{multirow}
\usepackage{orcidlink}
\journal{Physics Letters B}
\usepackage{ulem}

\newcommand{\stn}{s_{2n}}
\newcommand{\nucl}[2]{^{#1}\mathrm{#2}}
\renewcommand{\vec}[1]{\boldsymbol{#1}}

\biboptions{comma,sort&compress}

\begin{document}
	
	\begin{frontmatter}
		\title{Prediction of two-neutron halos in the $N=28$ isotones $\nucl{40}{Mg}$ and $\nucl{39}{Na}$}
		
		
\author[MCR,AKAL,RCNP,WARSAW]{Jagjit Singh\orcidlink{0000-0002-3198-4829}
\corref{mycorrespondingauthor}}
\cortext[mycorrespondingauthor]{Corresponding author}
\ead{jagjit.singh@manchester.ac.uk}
\author[SEVILLE]{J. Casal\orcidlink{0000-0002-5997-5860}}
\author[OMU,NITEP,RIKEN,HOKKAIDO]{W. Horiuchi\orcidlink{0000-0003-0039-3783}}
\author[MCR]{N. R. Walet\orcidlink{0000-0002-2061-5534}}
\author[WARSAW]{W. Satu{\l}a\orcidlink{0000-0003-0203-3773}}
\address[MCR]{Department of Physics and Astronomy, The University of Manchester, Manchester M13 9PL, UK}
\address[AKAL]{Department of Physics, Akal University, Talwandi Sabo, Bathinda, Punjab 151302, India}
\address[RCNP]{Research Centre for Nuclear Physics (RCNP), Osaka University, Ibaraki 567-0047, Japan}
\address[WARSAW]{Institute of Theoretical Physics, Faculty of Physics, University of Warsaw, ul. Pasteura 5, PL-02-093 Warsaw, Poland}
\address[SEVILLE]{Departamento de F\'{i}sica At\'{o}mica, Molecular y Nuclear, Facultad de F\'{i}sica, Universidad de Sevilla, Apartado 1065, E-41080 Sevilla, Spain}
\address[OMU]{Department of Physics, Osaka Metropolitan University, Osaka 558-8585, Japan}
\address[NITEP]{Nambu Yoichiro Institute of Theoretical and Experimental Physics (NITEP), Osaka Metropolitan University, Osaka 558-8585, Japan}
\address[RIKEN]{RIKEN Nishina Center, Wako 351-0198, Japan}
\address[HOKKAIDO]{Department of Physics, Hokkaido University, Sapporo 060-0810, Japan}
		
		
		\date{\today}

		\begin{abstract}
  The ground states of the nuclei $\nucl{40}{Mg}$ and $\nucl{39}{Na}$ are investigated using the hyperspherical formalism. Since they are located at the edge of the ``big island of inversion'', we concentrate on whether we are likely to find two-neutron Borromean halos in these nuclei. A three-body model with effective $n$-$n$ and ${\nucl{38}{Mg}}+n$ interactions is built for $\nucl{40}{Mg}$ based on the available data. We also give predictions for the low-lying spectrum of $\nucl{38}{Na} = {\nucl{37}{Na}}+n$ and two-neutron separation energy of the $\nucl{39}{Na}$ nucleus. 
  Depending on parameter choice, we report an increase in the matter radii in the range $0.1$\,-\,$0.5$~fm relative to those of the core nuclei. The results suggest a two-neutron halo structure in $\nucl{40}{Mg}$ for a subset of parameters, reinforcing the prediction of a Borromean halo nucleus. The calculations indicate that a two-neutron halo is even more likely for $\nucl{39}{Na}$. As expected, the halo is linked to the disappearance of the shell gap in these nuclei due to the inversion of the $2p_{3/2}$ and $1f_{7/2}$ orbitals. We study the total cross section for scattering of these nuclei from a carbon target using a Glauber model and show that these provide a clear signal to assess the halo structure.
		\end{abstract}

	\end{frontmatter}

	\section {Introduction}
	
The development of novel radioactive ion beam facilities has made it possible to study the shell evolution across the nuclear chart, and, most interestingly, near the neutron drip line. 
Neutron-rich nuclides display exotic structural features such as dampening of shell gaps \cite{Warburton1990,Dobaczewski1994,Sorlin2008}, shape coexistence \cite{Heyde2011, Otsuka2016, Nishibata2019} and formation of halos and deformed structure \cite{Tanihata1985, Motobayashi1995, Nakamura2009, Takechi2012, Hamamoto2012, Crawford2019, Bagachi2020}.
\begin{figure}[tb]
\centering
\includegraphics[width=0.9\linewidth]{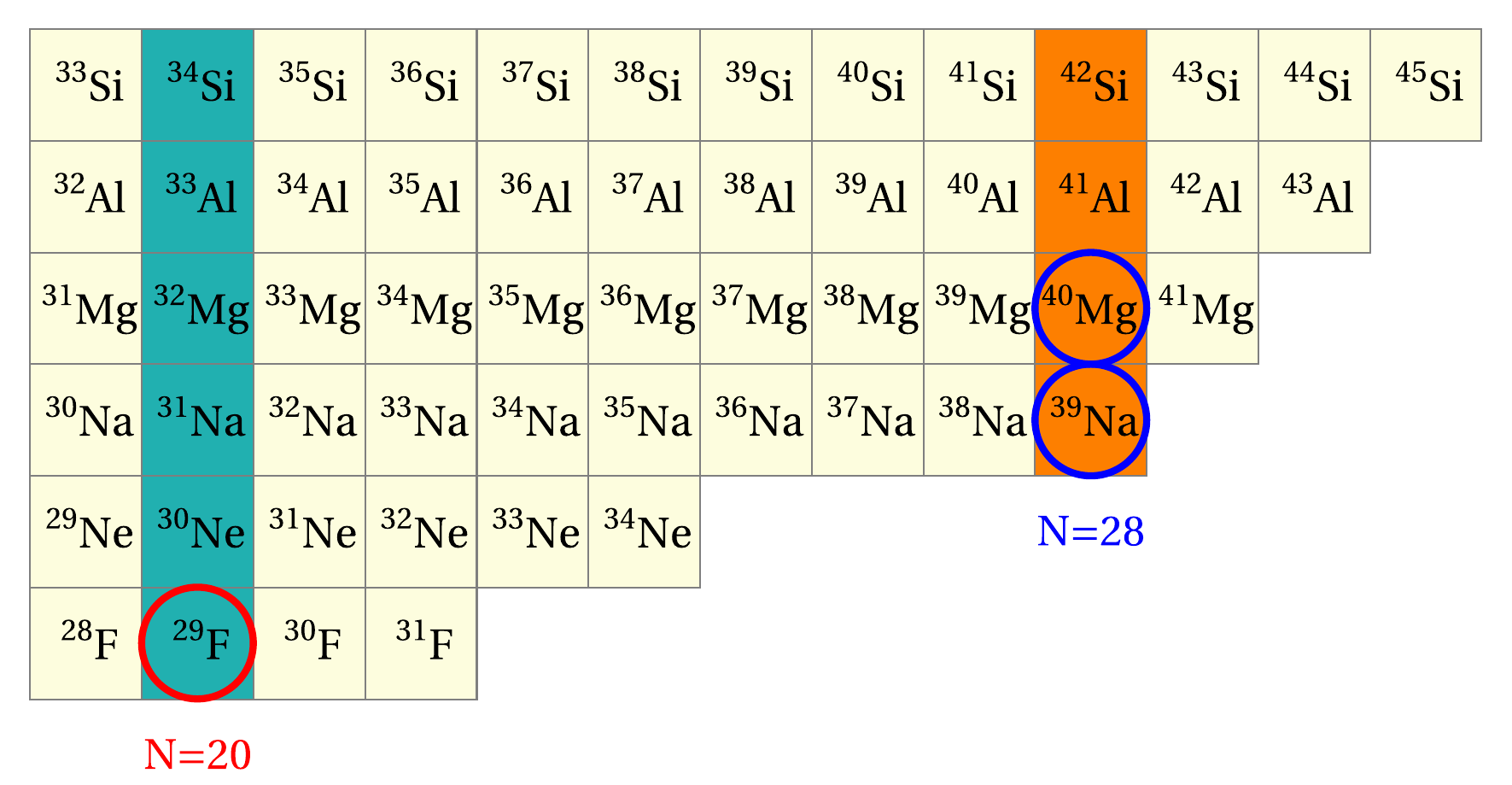}
\includegraphics[width=0.9\linewidth]{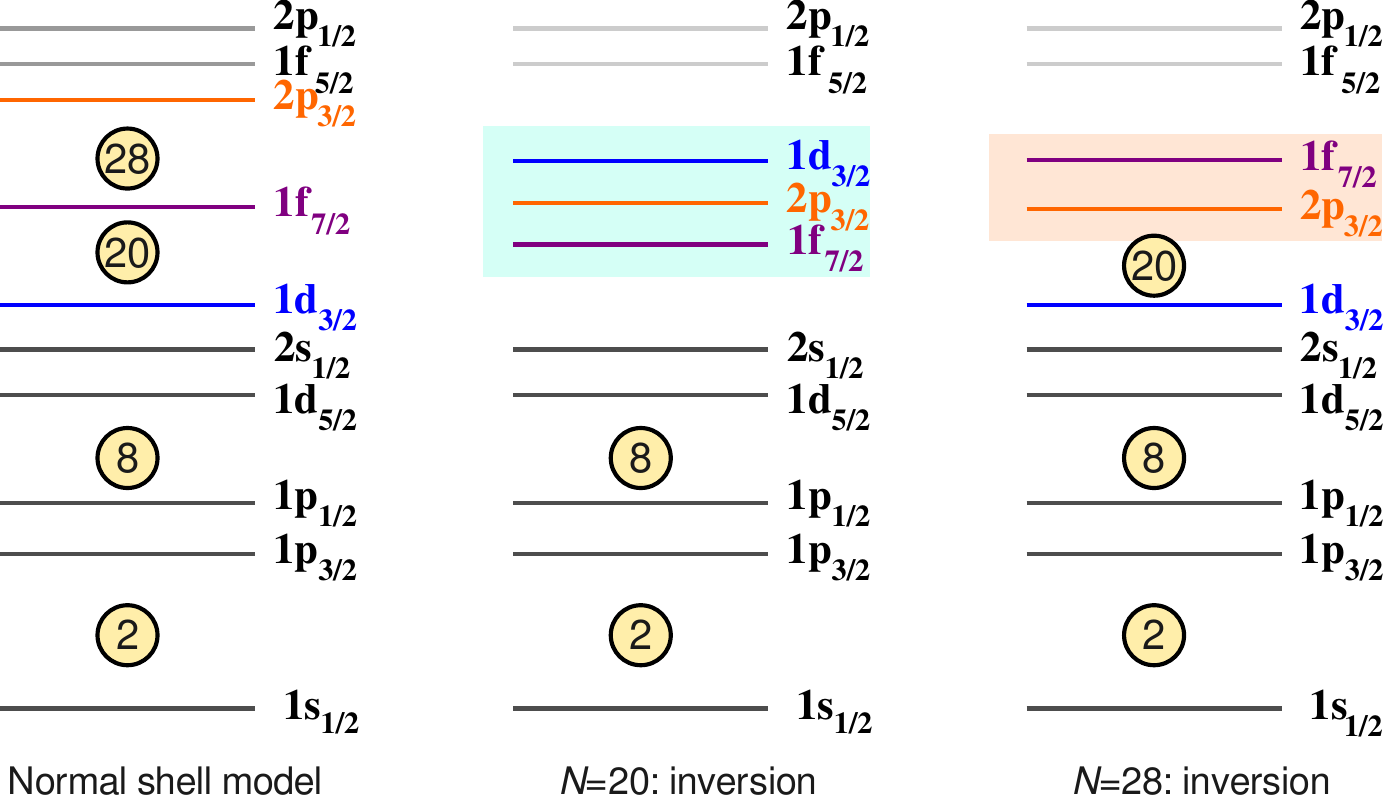}
\caption{(upper panel) A selected area of the nuclear Segr\`e chart for the $9 \le Z \le 14$ isotopes. The green and orange blocks correspond to neutron numbers $N=20$ and $28$, respectively. (lower panel) A sketch of the scenarios for the single-particle levels that lead to shell inversion. On the left, we show the shell gaps for stable nuclei (normal shell model),  and on the right, we show the active shells at $N=20$ and $N=28$ inside the coloured blocks. A change in the order of the shells leads to shell inversion.}
\label{FIG1}
\end{figure}

 Above $N=20$, an explanation of the magic numbers 
 requires a spin-orbit force \cite{Otsuka2020}. Thus, $N = 28$ is the smallest magic number that is sensitive to the spin-orbit force. As shown in the lower panel of Fig.~\ref{FIG1}, for $20\leq N\leq 28$ the $1f_{7/2}$ orbit is filled, while in the absence of spin-orbit splitting 
 the $1f_{7/2}$ and $2p_{3/2}$ orbits are almost degenerate. 
With modifications to the spin-orbit force, the spacing between these two orbits may change, and this can lead to an “inverted” order, where the $2p_{3/2}$ lies below the $1f_{7/2}$. The same result may also be caused by deformation \cite{Warburton1990,Horiuchi2022}. Such islands of inversion (IOI) have been found in isolated cases. It has also been shown that for $N = 20$ and $28$  the IOIs merge in the neon, sodium, and magnesium isotopic chains, creating a “big island of inversion”, called the B-IOI \cite{Caurier2014, Doornenbel2013}. 

The disappearance, or “melting”, of $N=28$ shell gap has been confirmed experimentally in neutron-rich Al, Si, S, and Ar isotopes \cite{Scheit1996, Glasmacher1997, Takeuchi2012, Soher2002, Doornenbel2013, Nowak2016, Calinescu2016, Crawford2019, Momiyama2020, Longfellow2020}. For $Z=11$, the lowest proton number in the B-IOI, the heaviest observed neutron-rich nucleus is $\nucl{39}{Na}$ \cite{Ahn2022, Zhang2023}, but the possibility of the existence of even heavier Na isotopes has not yet been excluded. Very few studies \cite{Zhang2023,Suzuki2021,Suzuki2022} examine the melting of $N = 28$ shell gap in Na and Mg isotones, among these only \cite{Zhang2023} delving into halo formation in $\nucl{39}{Na}$ using the deformed relativistic Hartree-Bogoliubov framework.

The recent observation of the disappearance of the $N=20$ shell gap at the low-$Z$ side of the $N=20$ chain (shown in the red circle in the upper panel of Fig.~\ref{FIG1}), has led to the identification of the $\nucl{29}{F}$ system as the heaviest known two-neutron Borromean-halo nucleus  \cite{Singh2020, Fortunato2020, Casal2020, Bagachi2020}. Motivated by this observation, it is interesting to explore the low-$Z$ side of the $N=28$ shell closure for potential two-neutron Borromean halos in the Na and Mg isotopes (as shown in the blue circles in the upper panel of Fig.~\ref{FIG1}). The magnesium isotopic chain also provides interesting candidates to explore the transition from strongly-bound to weakly-bound nuclei and the form of $n$-$n$ correlations at the limits of stability. Deformation will also play a role: e.g., the nucleus $\nucl{32}{Mg}$, with $N=20$,  can be described as a prolate-deformed rotor \cite{Motobayashi1995,Nowacki2021}. There is only limited knowledge of the properties of heavier neutron-rich Mg isotopes with $N\geq25$. The weakly-bound nucleus $\nucl{37}{Mg}$ ($N=25$) has been observed, and its ground state can be interpreted as a one-neutron halo with the valence neutron occupying $2p_{3/2}$ orbit \cite{Takechi2014, Kobayashi2014, Shubhchintak2015, Zhang202337Mg, Zhang202437Mg}. The neighbouring odd-A nucleus $\nucl{39}{Mg}$ was confirmed as unbound, indicating the important role of pair correlations.

For the present study, we start from the medium-mass open shell nucleus $\nucl{40}{Mg}$, which is the next two-neutron halo candidate after $\nucl{29}{F}$ according to various theoretical models ~\cite{Hamamoto2007,Caurier2014,Nakada2018,Macchiavelli2022}. The evaluated (\textit{i.e.}, not measured but estimated based on systematics) two-neutron separation energy of $\nucl{40}{Mg}$ is $s_{2n}=0.670\pm0.710\,\mathrm{MeV}$\cite{Wang2021}. This nucleus was first observed experimentally in 2007 \cite{Baumann2007} and more recently 
low-lying excited states of $\nucl{40}{Mg}$ have been studied using a one-proton removal reaction from $\nucl{41}{Al}$ \cite{Crawford2019}. Recent \textit{ab initio} calculations suggest large deformation in $^{40}$Mg~\cite{yuan2024}, though no connection to halo physics is made.

Because $\nucl{39}{Mg}$ is neutron unbound, it is reasonable to assume that the correlation between the two valence neutrons in $\nucl{40}{Mg}$ plays an important role in binding this system. Thus, the $\nucl{40}{Mg}$ nucleus, when interpreted as a $\nucl{38}{Mg}$ core plus two valence neutrons, provides an example of a Borromean system, similar to the well-studied two-neutron halo nuclei $^6$He \cite{Fort2014,SFV16EPJ} or $\nucl{11}{Li}$ \cite{Tanihata2013}. Because three-body models give a good description of
the structural properties of many Borromean nuclei, we apply the same technique, using effective three-body calculations to study the configuration mixing in and matter radius of the ground state of $\nucl{40}{Mg}$.  Given the uncertainties in the two-neutron separation energy, we explore how the allowed values influence the configuration mixing and matter radius, by analysing different scenarios. This allows us to explore the
possibility of halo formation in this region of the nuclear chart, well beyond the heaviest observed two-neutron halo in $\nucl{29}{F} $\cite{Singh2020, Fortunato2020, Casal2020, Bagachi2020}. The most recent atomic mass evaluation \cite{Wang2021} indicates that $\nucl{39}{Na}$ is unbound, contradicting Ahn \textit{et al.} \cite{Ahn2022} recent experimental discovery; our calculations within same three-body framework aim to predict the two-neutron separation energy ($\stn$) and matter radii for $\nucl{39}{Na}$ systematically. Finally, we also provide Glauber-model predictions for the total reaction cross sections of $\nucl{40}{Mg}$ and $\nucl{39}{Na}$ on a carbon target, which will further support the finding of halo formation in these systems. 
\section{Model Formulation} 
\label{sec:MF}
To focus on the effective $\text{core}+n+n$ system, we employ the hyperspherical formalism~\cite{Zhukov93,Nielsen01}. The eigenstates of the three-body Hamiltonian can be written as
\begin{equation}
  \Psi(\rho,\Omega) = \rho^{-5/2}\sum_{\beta}U_\beta(\rho)\mathcal{Y}_\beta(\Omega),
  \label{eq:3bwf}
\end{equation}
where $\rho^2=x^2+y^2$ is the hyperradius defined from the usual mass-scaled Jacobi coordinates $\left\{\boldsymbol{x},\boldsymbol{y}\right\}$, 
and the hyperangle $\alpha=\arctan{(x/y)}$ is contained in the angular variables $\Omega=\left\{\alpha,\widehat{x},\widehat{y}\right\}$. For simplicity, we choose $\boldsymbol{x}$ as the coordinate describing the neutron-neutron relative motion, and $\boldsymbol{y}$ as the coordinate between the core and the two neutrons. This is typically referred to as the Jacobi-$T$ representation. In Eq.~(\ref{eq:3bwf}), 
the index $\beta\equiv\{K,l_x,l_y,l,S_x\}j$ 
labels the different components of the wave function, with $\mathcal{Y}_{\beta}(\Omega)$ representing states of total angular momentum $j$ which follow the coupling order 
\begin{equation}
\mathcal{Y}_{\beta}(\Omega)=\left[\Upsilon_{Kl}^{l_xl_y}(\Omega)\otimes\kappa_{S_x}\right]_{j\mu}.
\label{eq:Upsilon}
\end{equation}
Here, $\Upsilon_{Klm_l}^{l_xl_y}(\Omega)$ are the hyperspherical harmonics~\cite{Zhukov93}, eigenstates of the hypermomentum operator $\widehat{K}$, and  $\kappa_{S_x}$ is a spin function. This equation implies that $\boldsymbol{l}=\boldsymbol{l}_x+\boldsymbol{l}_y$ is the total orbital angular momentum, $S_x$ is the coupled spin of the two neutrons, and the total $j$ results from $\boldsymbol{j}=\boldsymbol{l}+\boldsymbol{S}_x$. Note that we assume the core to be inert and its spin to be zero. More details can be found, for instance, in Ref.~\cite{Casal18}.

To determine the hyperradial functions $U_\beta(\rho)$ for bound states, we diagonalize the Hamiltonian in a suitable basis and focus on the negative-energy solutions. 
Thus, we expand
\begin{equation}
    R_\beta(\rho)=\sum_{i=0}^n C_{i\beta} U_{i\beta}(\rho),
    \label{eq:PS}
\end{equation}
where 
$C_{i\beta}$ are the diagonalization coefficients to be determined numerically. For this purpose, different bases have been used in various works~\cite{Desc03,Matsumoto04,MRoGa05}, and in the present work, we employ the analytical transformed harmonic oscillator (THO) basis from Ref.~\cite{Casal13}. The THO functions are built by performing a local scale transformation to the harmonic oscillator functions, changing their Gaussian asymptotic behaviour into an exponential decay, which is better suited to accurately describe bound nuclear states with a relatively small set of basis functions. This procedure was also adopted, for instance, in Refs.~\cite{Singh2020,Casal2020}.

The diagonalization of the three-body Hamiltonian requires the computation of the corresponding kinetic energy and potential matrix elements. With the above definition in hyperspherical coordinates, we can write~\cite{CasalTh,IJThompson04}
\begin{equation}
    T_\beta(\rho)=-\frac{\hbar^2}{2m} \left(\frac{d^2}{d\rho^2}-\frac{15/4+K(K+4)}{\rho^2} \right),
    \label{eq:top}
\end{equation}
for the kinetic energy operator, where 
$m$ 
is the mass of the nucleon. The coupling potentials are taken to be of the form
\begin{equation}
V_{\beta\beta'}(\rho)=\left\langle \mathcal{Y}_{\beta }(\Omega)\Big|V_{12}+V_{13}+V_{23} \Big|\mathcal{Y}_{\beta'}(\Omega) \right\rangle + \delta_{\beta\beta'}V_{\rm 3b}(\rho).
\label{eq:vcoupl}
\end{equation}
In this expression, $V_{ij}$ are the 
two-body interactions within the three-body composite system, 
to be fixed from the known information on the binary subsystems, and $V_{\rm 3b}(\rho)$ is a phenomenological three-body force, 
to account for effects not included in the two-body interactions alone~\cite{IJThompson04,MRoGa05,RdDiego10,Casal13}. This term contains the only free parameter in this model and is used to adjust the energy of the three-body states to the experimental values if known. 
\section{Two-body ($\text{core}+n$) subsystems}\label{sec:2bs}

The spectral properties of the $\text{core}+n$ subsystems play a key role in the study of the structure of Borromean three-body nuclei, as the two-body potentials are contained in the three-body Hamiltonian. In the present case, this amounts to fixing a phenomenological $\nucl{38}{Mg}+n$ and $\nucl{37}{Na}+n$ potential to describe the low-lying continuum spectrum of $\nucl{39}{Mg}$ and $\nucl{38}{Na}$, respectively. 

We will start with $\nucl{38}{Mg}+n$ system. Though $N=26$ has a partially filled $1f_{7/2}$ subshell, we model $\nucl{39}{Mg}$ as a $\nucl{38}{Mg}$ core surrounded by an unbound neutron moving in the $f_{7/2}$, $p_{3/2}$, $f_{5/2}$, and $p_{1/2}$ orbitals in a simple independent-particle shell-model picture. Even though the distinction between the core and valence neutrons is not completely obvious, for simplicity, we nevertheless make an inert-core approximation. This means that the effects from internal rearrangement or core-valence exchange must be contained in effective potential parameters. Note that a similar approach has been successfully followed in other three-body calculations, such as the $^{27}\text{F}+n$ and the $^{29}\text{F}+n$ subsystems for the description of $\nucl{29}{F}~$\cite{Singh2020, Fortunato2020, Casal2020} and $\nucl{31}{F}~$\cite{GSingh2022}. Core excitations could be incorporated by using, for instance, a particle-rotor model~\cite{Macchiavelli2022} with an effective core deformation parameter.

The only theoretical study on the spectrum of the neutron-unbound nucleus $\nucl{39}{Mg}$ reports an indication of a low-lying resonance at $0.129$\,MeV with $J^\pi$ either $7/2^-$ or $3/2^-$ \cite{Fossez2016}. However, due to the unavailability of experimental data, this does not rule out the existence of other possibilities such as a single resonant structure or more than two states, and in any case, the spin-parity assignment is not unambiguous.
\begin{table}[tb]
\caption{Parameter sets for the $\nucl{38}{Mg}+n$ (upper-panel) and $\nucl{37}{Na}+n$ (lower-panel) Woods-Saxon interactions, Eq.~(\ref{eq:WS}). Here $a$ is diffuseness, $V^{(l)}_0$ is the potential depth and $E_{R}$ is the position of the resonances. Note that $r_0=1.25$~fm,  and $V_{ls}=16.842$~MeV (for $\nucl{38}{Mg}+n$) and $16.324$~MeV (for $\nucl{37}{Na}+n$) are fixed.}
\vspace{0.2cm}
\small
\centering
\begin{tabular}{lccccccr}
\toprule\toprule
 &  &  &  $\nucl{38}{Mg}+n$      & &  \\ [1ex]
\toprule\toprule
Set & Scenario & $a$\,(fm)  &  $lj$      & $V^{(l)}_0$\,(MeV) & $E_{R}$\,(MeV) \\ [1ex]
\toprule
\textbf{1}& Normal        & 0.70 & $f_{7/2}$ & 38.225   & 0.129    \\
          &  &      & $p_{3/2}$ & 38.225   & 0.349    \\ [1ex]
\textbf{2}& Degenerate   & 0.75 & $f_{7/2}$ & 38.400   & 0.129    \\
          &                &      & $p_{3/2}$ & 38.400   & 0.135    \\ [1ex]
\textbf{3}& Inverted     & 0.75 & $f_{7/2}$ & 37.880   & 0.349    \\
           & &      & $p_{3/2}$ & 38.425   & 0.130    \\ [1ex]
\toprule\toprule
 &  &  &  $\nucl{37}{Na}+n$     & &  \\ [1ex]
\toprule\toprule
\textbf{1}& Degenerate       & 0.70 & $f_{7/2}$ & 38.225   & 0.539    \\
          &  &      & $p_{3/2}$ & 38.225   & 0.599    \\ [1ex]
\textbf{2}& Inverted       & 0.75 & $f_{7/2}$ & 38.400   & 0.522    \\
          &  &      & $p_{3/2}$ & 38.400   & 0.271    \\ [1ex]
\textbf{3}& Inverted      & 0.75 & $f_{7/2}$ & 37.880   & 0.734    \\
           &                &      & $p_{3/2}$ & 38.425   & 0.265     \\ [1ex]
\toprule\toprule
\end{tabular}
\label{T1}
\end{table}

Considering the limited information available, we model the $^{38}\text{Mg}+n$ interaction as a Woods-Saxon potential including only central and spin-orbit terms,
\begin{equation}
    V_{{\rm core}+n}(r) = \left(-V^{(l)}_0+V_{ls}\lambda_\pi^2 \vec{l}\cdot\vec{s}\frac{1}{r}\frac{d}{dr}\right)\frac{1}{1+\exp[(r-R_c)/a]}\,,\label{eq:WS}
\end{equation} where $V^{l}_0$ is in general $l$-dependent, and $R_c=r_0A_c^{1/3}$  with $A_c$  the mass number of the core. The spin-orbit interaction is written in terms of the Compton wavelength $\lambda_\pi=1.414$~fm. Following Ref.~\cite{HOR10}, the spin-orbit strength is taken to follow the systematic trend \cite{BOHR} and has the value $V_{ls}=16.842$~MeV for $\nucl{38}{Mg}+n$ and $16.324$~MeV for $\nucl{37}{Na}+n$. The value $r_0=1.25$~fm is as originally suggested for $\nucl{31}{Ne}$ ~\cite{HOR10,Singh2020}, and we examine three scenarios, see Table~\ref{T1}. In Set $1$ and $2$, $V_0$ is chosen to be $l$-independent. In set $1$ it is adjusted to fix the $f_{7/2}$ ground-state resonance of $\nucl{39}{Mg}$ at $0.129$\,MeV, corresponding to the prediction of Ref.~\cite{Fossez2016}. This set gives a \textit{``normal"} shell-model scenario, with an additional $p_{3/2}$ resonance appearing at higher energy, $0.349$\,MeV.  

To explore the competition of the $p$-wave with the $f$-wave ground state, we consider two additional scenarios defined by Sets $2$ and $3$, respectively. In the \textit{``degenerate"} case, we tune the $f$-wave strength and diffuseness parameter $a$ so that the $f_{7/2}$ and $p_{3/2}$ resonances become nearly degenerate. The \textit{``inverted"} scenario assumes the ground state of $\nucl{39}{Mg}$ to be the $p$-wave resonance. Here we need to allow orbital dependence, and we have different potential depths for $l=1$ and $l=3$.

These potentials generate Pauli forbidden single-particle states, which would result in unphysical eigenstates of the three-body Hamiltonian, and need to be removed. In this work, this is achieved through a supersymmetric transformation~\cite{Baye87, Baye287}.

As we do not have either theoretical or experimental predictions and data for $\nucl{38}{Na}$, we use the same potential parameters as for $\nucl{39}{Mg}$. The only changes are $R_c$ and the spin-orbit strength, as explained above. The latter takes into account effectively the tensor force between the unpaired proton and valence neutron~\cite{HAG,Otsuka2010}. Indeed, since $\nucl{39}{Na}$ has an odd proton number, its ground state angular momentum is not zero. However, we have found before \cite{Singh2020, Casal2020B19, GSingh2022} that approximating the ground state of these light systems as a $0^+$ state gives reasonable results. The lower half of Table~\ref{T1} shows the parameter sets for the $\nucl{37}{Na}+n$ potential. Our predictions for the low-lying spectrum of $\nucl{38}{Na}$ show no ``normal'' ordering of the single-particle states; most scenarios present an inversion, indicating a shell-evolution effect, that we describe effectively by just changing the spin-orbit strength.

\section{Three-body ($\text{core}+n+n$) calculations}
\label{sec:3bc}
Using the core$+n$ potentials outlined in the preceding section, we compute the three-body ground-state of the $\nucl{40}{Mg}$ and $\nucl{39}{Na}$ nuclei assuming a simple description in which two valence neutrons couple to $J^\pi=0^+$. To provide a more comprehensive description, it is essential to have precise experimental data on the core$+n$ spectrum and a thorough understanding of the intricate core$+n$ interactions that can result in the splitting of single-particle energy levels. In the present calculation, we also need a prescription for the $nn$ interaction. We use the Gogny-Pires-Tourreil (GPT) potential~\cite{Gogny70} as in previous works~\cite{Lovell17,IJThompson04,Casal13}. 
It is important to emphasize that when it comes to the ground-state properties of $\text{core}+n+n$ systems, the specific selection of the $nn$ interaction is not particularly critical, as long as the chosen interaction adequately represents $nn$ scattering data, as discussed in Ref.~\cite{Zhukov93}. As previously introduced in Eq.~(\ref{eq:vcoupl}), we incorporate a phenomenological three-body force to obtain the energy of the three-body ground state, following a methodology similar to that employed in our earlier studies \cite{Casal13,Singh2020,GSingh2022}. The three-body problem is solved by diagonalizing the Hamiltonian in a THO basis including hyperspherical harmonics up to a maximum hypermomentum $K_{max}$ in the wave-function expansion (Eq.~(\ref{eq:3bwf})), and $i=0,\dots,N$ excitations for the hyperradial functions (Eq.~(\ref{eq:PS})). Note that fixing $K_{max}$ implies that the orbital angular momenta associated with each Jacobi coordinate are restricted to $l_x+l_y \le K$~\cite{Zhukov93}. In this work, $K_{max}=46$ and $N=30$ are found to provide converged results.

In order to compute the $\nucl{40}{Mg}$ and $\nucl{39}{Na}$ matter radius within the present three-body model~\cite{CasalTh} using $R_m = \sqrt{\frac{1}{A}\left(A_c R_c^2 + \langle r_{nn}^2\rangle\right)}$, the size of the $\nucl{38}{Mg}$ and $\nucl{37}{Na}$ cores are required as input. We have used $R_c=R_m(\nucl{38}{Mg},\nucl{37}{Na})=3.600, 3.637$~fm, respectively, adopted from Ref.~\cite{Watanabe2014,GENG2004}. These are somewhat uncertain, so it is more precise to look at a comparison of the differences across various potential models as well as the relative increase between core and three-body systems.

\begin{figure}[tb]
\centering
\includegraphics[width=0.9\linewidth]{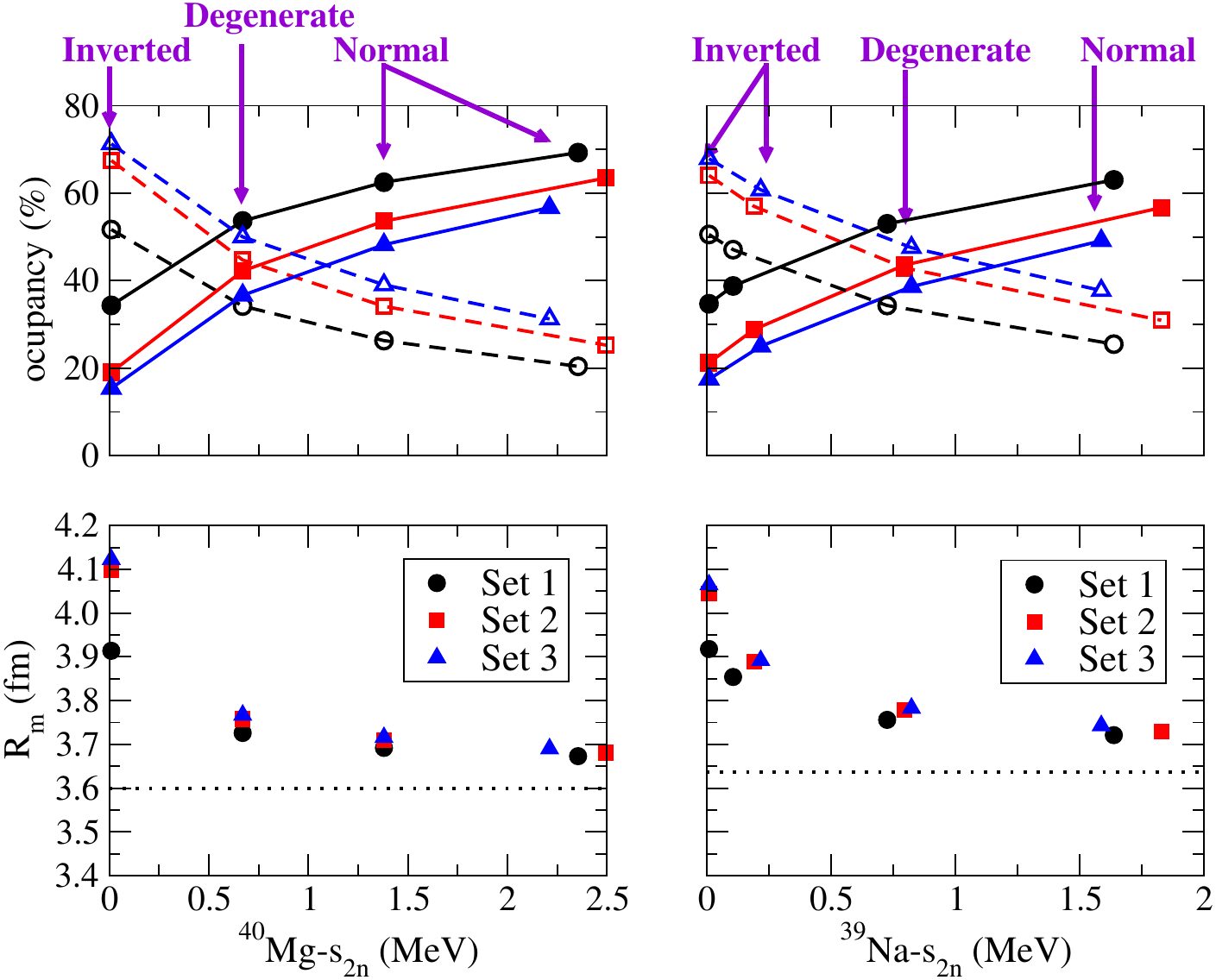}
\caption{(upper panel) The occupancy of the $(f_{7/2})^2$ (solid lines) and $(p_{3/2})^2$ orbits (dashed lines) for different parameter sets in $\nucl{40}{Mg}$ (left) and $\nucl{39}{Na}$ (right), respectively, as a function of $\stn$. (lower panel) Variation of matter radii (R$_m$) for $\nucl{40}{Mg}$ and $\nucl{39}{Na}$ with $\stn$. The black dotted line shows the core radius,  $\nucl{38}{Mg}$ and $\nucl{37}{Na}$, respectively. It should be noted that the labels Inverted, Degenerate and Normal \textit{pictures} here refer to the outcome of full three-body calculations and should not be confused with the single-particle \textit{scenarios} of Table 1.}
\label{FIG2}
\end{figure}

With all these ingredients, our three-body results for the ground state of $\nucl{40}{Mg}$ and $\nucl{39}{Na}$ using the scenarios for the $\text{core}+n$ potentials are shown in Fig~\ref{FIG2}. As already discussed, there are large uncertainties in the evaluated $\stn$ value \cite{Wang2021} of $\nucl{40}{Mg}$ and no information is available for the $\stn$ value of $\nucl{39}{Na}$. Given these uncertainties, we explore the sensitivity of the configuration mixing and matter radius of the ground state with $\stn$, using different values of the three-body potential strength. For $\nucl{40}{Mg}$, $V_{3b}=0$ leads to an overbinding of the system, so we need to choose $V_{3b}>0$ to get $\stn$ corresponding to the lower, central and upper limits of the evaluation ($0.010$, $0.670$ and $1.380$ MeV). Note that 
the lower limit would result in an unbound state. However, the experimental findings support a bound ground state, so we have considered a barely bound case as well. For $\nucl{39}{Na}$, due to the lack of experimental information and systematic information on $\stn$, it seems reasonable to keep the range of $V_{3b}$ values determined for $\nucl{40}{Mg}$ and make predictions. With this prescription, the shallowest case becomes unbound for all sets. To restrict the situation to bound states, we modify $V_{3b}$ in that case so that $\stn$ is $0.010\,\text{MeV}$, since we are interested in giving predictions for a bound halo nucleus, which could be tested experimentally. We predict $\stn$ for $\nucl{39}{Na}$ between $0.010$-$0.824$ ($1.828$)\,MeV without (with) $V_{3b}=0$ cases.

The computed partial wave content and radii for each $\stn$ value are shown in the upper panel and lower panel, respectively, of Fig~\ref{FIG2}. 
As expected, shallower (bound) ground states yield larger $p$-wave occupancy which in turn leads to an ``inverted'' picture, in contrast to the deeply-bound ground states which lead to the ``degenerate'' and ``normal'' pictures for all sets. As can be seen from the upper panel of Fig.~\ref{FIG2}, when moving from Set 1 to Set 3 the $p$-wave content in the ground state increases, whereas the $f$-wave content decreases for each choice of $\stn$ value. We remark that the similar behavior was found in Ref.\cite{Masui2020}. 

Similarly, we see in the lower panel of Fig.~\ref{FIG2} that, as expected, the shallowest cases give larger matter radii in contrast with more deeply bound ground states, which result in smaller matter radii for all sets in both systems considered. More interestingly, we note that the effect of the value of $\stn$ on the change in radius is most substantial when we have a shell inversion. However, due to the rather heavy cores, the calculated total radii are similar for the scenarios considered. Even with a substantial neutron halo, for a fixed $\stn$ the variations amount up to a maximum of $5\,\%$ difference for both systems. Differentiating scenarios solely by radius could pose experimental difficulties. Additionally, data on knockout or transfer reactions that can reveal the partial wave content in the ground state of these nuclei would be very useful to tighten the constraints on theoretical models and enable us to distinguish between various wave functions that have been presented in this text.

In the lower panel of Fig~\ref{FIG2}, we show the radius of the $\nucl{38}{Mg}$ and $\nucl{37}{Na}$ cores as a horizontal dotted black line which makes it easy to extract the relative change in matter radius of $\nucl{40}{Mg}$ and $\nucl{39}{Na}$ relative to the radius of the appropriate core, $\Delta R = R_m-R_{core}$. With the current uncertainties in $\stn$ and the low-lying spectrum of subsystems, the difference of the matter radius of these nuclei and their cores ranges between $0.1$-$0.5$\,fm for different choices of potential sets and $\stn$ values. This number is notably smaller than those for well-established halo nuclei such as $\nucl{6}{He}$ or $\nucl{11}{Li}$, but is close to the heaviest known two-neutron halo, $\nucl{29}{F}$, which is only $0.35$\,fm larger than the $\nucl{27}{F}$ core 
\cite{Singh2020,Casal2020,Bagachi2020}. It is worth mentioning that, recently, we have shown that the halo formation in $\nucl{29}{F}$ is connected to the weakening of the $N=20$ shell gap, which leads to the intrusion of the $p_{3/2}$ orbital and reduced binding \cite{Singh2020,Fortunato2020,Casal2020}, which leads to halo formation. The same situation is observed when $N=28$ melts, with the difference that the $p_{3/2}$ orbital is within the same major shell. 
As depicted in Fig.\Ref{FIG2}, our three-body findings reveal a substantial increase in $\Delta R$ for the least deeply bound state, with a substantial $p_{3/2}$ component. This suggests that our study provides evidence for a halo structure in the ground states of both $\nucl{40}{Mg}$ and $\nucl{39}{Na}$. However, experimental cross section measurements are necessary to confirm this assertion.

\begin{figure}[tb]
\centering
\includegraphics[width=0.9\linewidth]{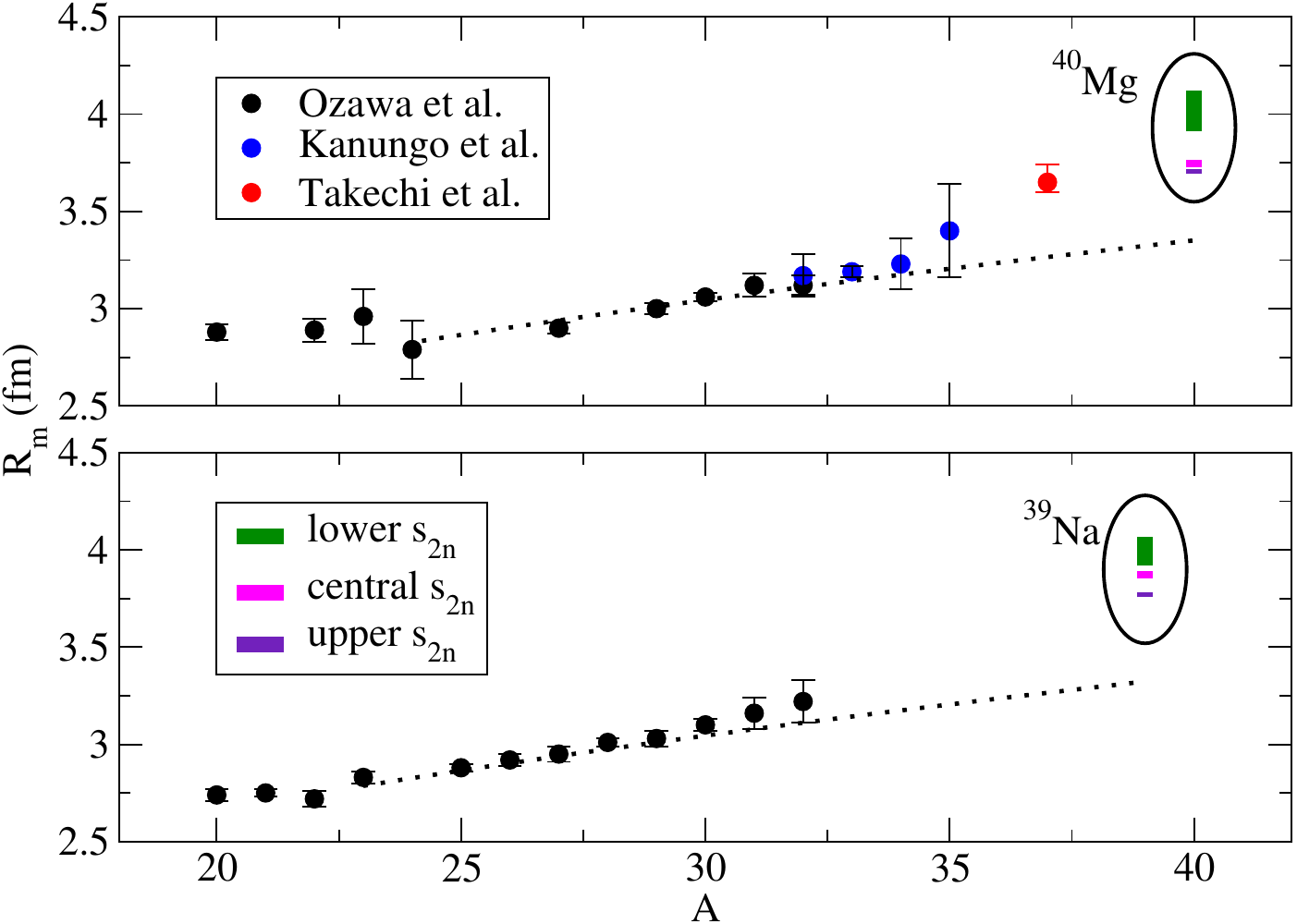}
\caption{The Variation of matter radii ($R_m$) for Mg (upper panel) and Na (lower panel) isotopes with mass number $A$. Experimental values, shown in black, blue, and red circles, were taken from Refs. \cite{Ozawa01}, \cite{Kanungo2011}, and \cite{Takechi2014}, respectively. The dotted black line corresponds to the $R_0A^{1/3}$ fit for $A=24$-$37$ for Mg-isotopes and $A=23$-$32$ for Na-isotopes. It should be noted that each colour band associated with a specific $\stn$ corresponds to the range of values for Sets $1$, $2$, and $3$, and the same applies to Fig. ~\ref{FIG5}.} 
\label{FIG2A}
\end{figure}

In Fig.~\ref{FIG2A} we show the variation of matter radii for the Mg and Na isotopes with increasing mass numbers. The experimental derived values for the radii for $\nucl{20-32}{Mg}$ and $\nucl{20-32}{Na}$ (shown in black circles) are adopted from Ref.~\cite{Ozawa01}, while those for $\nucl{32-25}{Mg}$ are taken from Ref.~\cite{Kanungo2011} (shown in blue circles). Data for the one-neutron halo $\nucl{37}{Mg}$ (shown as a red circle) is taken from Ref.~\cite{Takechi2014}. The dotted black lines in the figure correspond to a weighted fit of the experimental data points of the form $R_0A^{1/3}$ for $A=24$–$37$ for Mg isotopes and $A=23$–$32$ for Na isotopes, respectively. It must be noted that the sharp rise of radius in $\nucl{37}{Mg}$ compared to the nearest lighter isotope can be attributed to the fact that $\nucl{37}{Mg}$ is itself a one-neutron halo nucleus \cite{Takechi2014}.
It can be easily seen from Fig.~\ref{FIG2A} that the radii of $\nucl{40}{Mg}$ and $\nucl{39}{Na}$ are larger than the standard $R_0A^{1/3}$ fitted value for all choices of $\stn$. The present calculations provide evidence of a modest halo in all instances, with a more pronounced effect observed as the occupation of the $p_{3/2}$ intruder orbital increases, akin to the case of $\nucl{29}{F}$. Thus, this observation implies a likely two-neutron halo structure in the ground state of $\nucl{40}{Mg}$ and $\nucl{39}{Na}$, and the corresponding melting of the traditional $N=28$ shell gap is due to the intrusion
of the $p_{3/2}$ orbital. 

\begin{figure}[tb]
\centering
\includegraphics[clip,width = 3.2in]{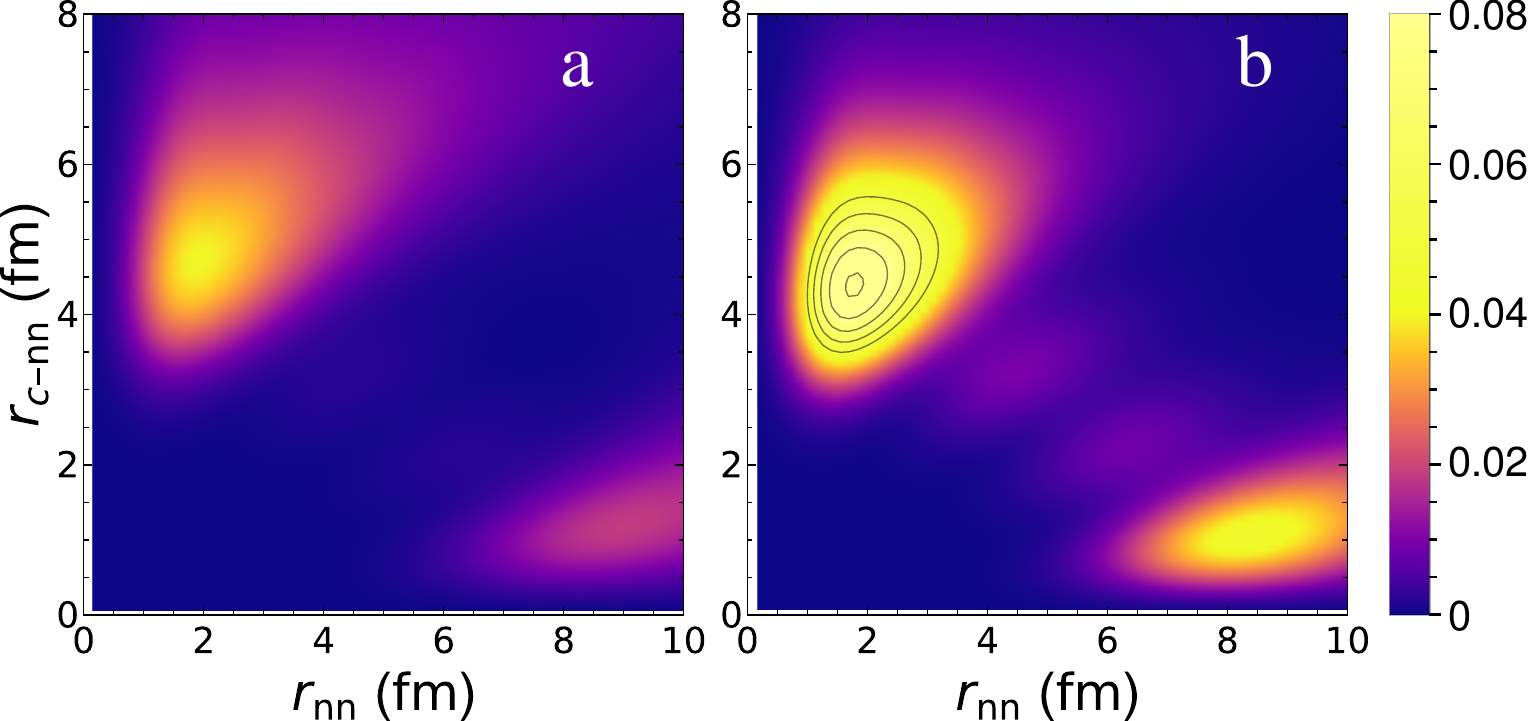} 
\caption{The ground state probability density distribution (in units of $\mathrm{fm}^{-2}$) for $\nucl{40}{Mg}$ for Set-3 (a) with $\stn=0.010$ MeV and (b) with $\stn=1.380$ MeV. The contour lines in (b) show the shape of the peak above a probability of $0.04$, which is twice as high as that in (a).}
\label{FIG3}
\end{figure}
We can gain a more in-depth understanding of the wave function by examining the probability densities in the Jacobi-T system, scaling $x$ and $y$ to the distance between the valence neutrons ($r_{nn}$) and that between the center of mass of the neutrons and the core ($r_{c-nn}$). Figure~\ref{FIG3} illustrates density distributions for $\nucl{40}{Mg}$ in the inverted scenario, considering both lower and upper limits of $\stn$, which encourages halo formation in its ground state. The density distributions for $\nucl{39}{Na}$ closely resemble those of $\nucl{40}{Mg}$ and are consequently not shown. Both selections of the $\stn$ result in a peak where two neutrons are situated near each other at a distance from the core, the di-neutron peak. However, the latter choice leads to a more concentrated wave function, which is a consequence of the larger $f$-wave component (evidenced by the presence of four peaks in Fig.~\ref{FIG3}b). It is also evident that the valence neutrons explore shorter relative distances. These characteristics are reflected in the root-mean-square (rms) values of the coordinates, with values of $11.35$ and $8.06$\,fm for $r_{nn}$ and $r_{c-nn}$, respectively, in the shallow case, and $6.43$ and $4.55$\,fm in the deeper case. 

In summary, the results presented in this section provide evidence for the formation of halos in the ground state of both $\nucl{40}{Mg}$ and $\nucl{39}{Na}$ systems. The formation of these halos is attributed to the dissolution of the $N=28$ shell gap, leading to the intrusion of the $p_{3/2}$ orbital and subsequent weak binding.

\section {Total reaction cross section}\label{sec:xsec}
Experimentally, a very obvious way to determine whether a nucleus is a halo nucleus, is to look for an enhanced reaction cross section. Thus we examine the total reaction cross section ($\sigma_R$) by employing the conventional Glauber theory~\cite{Glauber}. We utilize the nucleon-target formalism~\cite{AbuIbrahim00} and apply the nucleon-nucleon profile function from Refs.~\cite{AbuIbrahim08, AbuIbrahim09E, AbuIbrahim10E}.This approach has proven effective in various high-energy nucleus-nucleus collision reactions, particularly those with unstable nuclei, and it successfully replicated isotope dependence of the total reaction cross sections with appropriate density distributions~\cite{HOR10,Horiuchi2006,Horiuchi2007a,Horiuchi2007b,Horiuchi12,Horiuchi15jps,Nagaisha2018}. 
\begin{figure}[tb]
\centering
\includegraphics[width=0.9\linewidth]{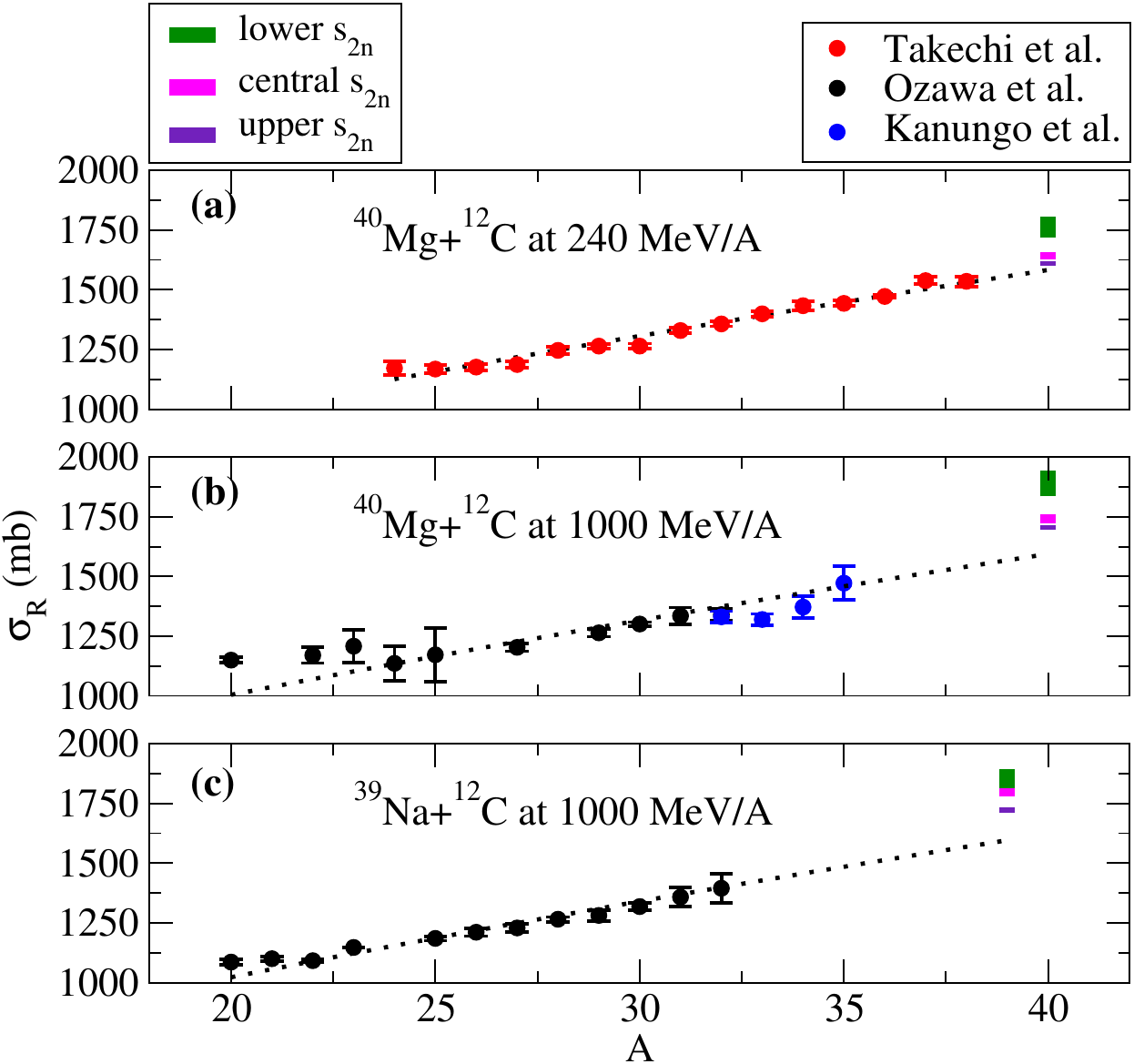}
\caption{Measured and calculated total reaction cross section for different isotopes of Mg and Na at different energies. Theoretical results (shown with green, magenta, and indigo bars) for $\nucl{40}{Mg}$$+$$\nucl{12}{C}$ are shown at (a) $240$\,MeV/A and (b) $1000$\,MeV/A. Experimental data shown with red, blue, and black dots are taken from \cite{Takechi2014} (at $240$\,MeV/A), \cite{Kanungo2011} (at $900$\,MeV/A), and \cite{Ozawa01} (at $950$\,MeV/A), respectively. (c) Shows the results for $\nucl{39}{Na}$$+$$\nucl{12}{C}$ at $1000$\,MeV/A and experimental data shown with black dots is taken from \cite{Ozawa01} (at $950$\,MeV/A). The dotted black line corresponds to a $\sigma_0A^{2/3}$ fit to guide the eye of the reader.}
\label{FIG5}
\end{figure}

The additional theoretical inputs required for this reaction model include the density distributions of both the projectile and target nuclei. In the case of $\nucl{40}{Mg}$, we first generate a harmonic-oscillator (HO) type density distribution for $\nucl{38}{Mg}$ that reproduces the measured total reaction cross section of $\nucl{38}{Mg}$$+$$\nucl{12}{C}$ by Takechi \textit{et al.,} \cite{Takechi2014} at $240$MeV/nucleon ($1535\pm 21$ mb) by assuming a simple shell model state $(1f_{7/2})_\nu^8\otimes(1d_{5/2})_\pi^4$.  
The density for $\nucl{40}{Mg}$ was constructed by simply adding the $2n$ densities calculated in the previous section to the $\nucl{38}{Mg}$ density. No center-of-mass correction was applied in these calculations, but we believe this is a safe strategy for a system with such a heavy core. Using this prescription, we predict the $\sigma_R$ for $\nucl{40}{Mg}$ at different incident energies, $240$ MeV/nucleon (lies between $1601$-$1807$ mb, shown in Fig.~\ref{FIG5}(a)) and $1000$ MeV/A (lies between $1695$-$1944$ mb, shown in the Fig.~\ref{FIG5}(b)). As can be seen from Fig.~\ref{FIG5}(a) and (b), the predicted values of $\sigma_R$ for $\nucl{40}{Mg}$ show significant enhancement with respect to the observed $\sigma_R$ in the lower-$A$ isotopes for both choices of energy. 
This shows that we can use these reaction cross sections as an indication of the melting of the $N=28$ shell closure.

We use the same HO parameterization established for $\nucl{38}{Mg}$ for $\nucl{37}{Na}$. 
Hence, we forecast the $\sigma_R$ for $\nucl{39}{Na}$ at an energy of $1000$\,MeV/nucleon, as depicted in Fig.~\ref{FIG5}(c) (lies between $1710$-$1894$ mb). Our computed estimates of $\sigma_R$ for $\nucl{39}{Na}$ reveal a noteworthy increase compared to the observed $\sigma_R$ in its lower-mass isotopes. This reinforces our earlier finding in the $\nucl{40}{Mg}$, which points to the formation of a two-neutron halo in the ground state of $\nucl{39}{Na}$, leading to the disruption of the neutron shell at $N=28$.

\section{Conclusions}\label{sec:conc}

We have analysed the configuration mixing and matter radius of the ground states of the nuclei $\nucl{40}{Mg}$ and $\nucl{39}{Na}$ at the potential $N=28$ shell closure. Since we are using an effective three-body model, we can study two-neutron halo formation. Unfortunately, there is very little information on the properties of $\nucl{39}{Mg}$, which we would need to constrain our models. We have instead analysed three scenarios that span the range of allowed parameters. We have then used the same parameters for both $\nucl{40}{Mg}$ and $\nucl{39}{Na}$. We have thus mapped out the whole range of the expected $\stn$ values in $\nucl{40}{Mg}$ and   $\nucl{39}{Na}$. 

Our results indicate a picture with mild to strong mixing of the intruder $p_{3/2}$ orbit with the normal $f_{7/2}$ orbit due to the magnitude of the $\text{core}+n$ potentials in our three scenarios. This seems difficult to avoid in $\nucl{39}{Na}$, where most scenarios lead to a shell inversion. It seems only plausible for $\nucl{40}{Mg}$, where a normal ordering is also possible. The mixing enhances the di-neutron configuration in the ground-state density. Due to the heavy masses of these nuclei, the computed matter radii vary only up to $5$\%  for a fixed $\stn$ and considering the different choices of the core$+n$ interaction. This makes it challenging to verify this experimentally, since one must make highly accurate mass measurements. On the theory side, we need a thorough understanding of the low-energy continuum spectrum of core$+n$ subsystems. Additionally, the inclusion of transfer or knockout data, capable of probing the partial-wave characteristics, would be highly valuable in distinguishing between the scenarios used in this study.

The matter radii of $\nucl{40}{Mg}$ and $\nucl{39}{Na}$ show an increase of $0.1$ to $0.5$\,fm compared to their respective cores $\nucl{38}{Mg}$ and $\nucl{37}{Na}$, lending support to the idea of a potential halo structure in their ground state.
This assertion gains further backing from the elevated predicted reaction cross sections for these nuclei in comparison to the lower mass isotopes. Nevertheless, it is imperative to verify this conclusion through experimental measurements of interaction cross sections. 
Future perspectives include incorporating core deformation within the effective three-body description explicitly, to assess its influence on the halo structure in this region of the nuclear chart.
\section*{Acknowledgments}
We thank L. Fortunato for careful reading of the manuscript. This work was supported by the UK Science and Technology Funding Council [grant number ST/V001116/1] (JS and NRW), by the Polish National Science Centre (NCN) [under Contract No 2018/31/B/ST2/02220] (JS and WS), by the Spanish MCIN/AEI/10.13039/501100011033 [under I+D+i project No. PID2020-114687GB-I00] (JC), and by the JSPS KAKENHI
Grants Nos. 18K03635 and 22H01214 (WH). 
\bibliography{ref}

\end{document}